\documentstyle[aps,pra,epsf]{revtex}
\begin{document}
\draft
\tighten
\twocolumn[\hsize\textwidth\columnwidth\hsize\csname
@twocolumnfalse\endcsname
\title{Gate errors in solid state quantum computation}
\author{Xuedong Hu and S. Das Sarma}
\address{Condensed Matter Theory Center, Department of Physics,
University of Maryland, College Park, MD 20742-4111}
\maketitle
\begin{abstract}
We review our work on the interplay between non-resonant gates
and solid state environment in various solid state quantum computer
architectures and the resulting gate errors.  Particular, we show that
adiabatic condition can be satisfied in small quantum dots, while 
higher energy excited states can play important role in the evolution
of a Cooper-pair-box based quantum computer model.  We also show that
complicated bandstructure such as that of Si can pose a severe gate
control problem.
\end{abstract}
\vskip2pc]
\narrowtext

\section{Introduction}

It has been pointed out for a long time that quantum mechanics 
may provide great advantages over classical physics in physical
computation \cite{Deutsch,Feynman}.  The recent rapid growth
of research on quantum computation \cite{Review} started after 
Shor's factorization algorithm \cite{Shor1} and quantum
error correction codes \cite{Shor2,Steane} were developed.  
For a quantum system to be used as a quantum computer (QC), it has to
satisfy some stringent conditions \cite{DiVincenzo}.  In short, it
should possess a scalable Hilbert space; the state of such a system
should be easily initialized; the system should have a long
deocherence time; there should a set of universal unitary gates
applicable to the system; and last but not least, every single
quantum bit (qubit) of the system should be faithfully measured.  Here
we would like to review our work on the quantum gates and their
operations in various solid state quantum computer architectures.

Many two-level systems have been proposed as candidates for qubits
in a solid state quantum computer.  Typical examples include electron
spins, nuclear spins, electron charge states, Cooper pair charge
states, superconducting flux states, and many more
\cite{LD,HD,Vrijen,Kane,Privman,Schon,Averin}.  One major
motivation for these solid state devices is their potential in
scalability.  However, solid state structures also present complex
environments and  fast decoherence rates \cite{HSD}. Furthermore, in
most solid state QC schemes, non-resonant gate operations
\cite{LD,Kane,Schon,Nakamura} are crucial or important components. 
It is thus necessary to understand how the environmental elements
affect the QC coherent evolution, and clarify the effects of
imperfections in non-resonant gate operations.  In the following, we
will review some of the results we have obtained for three different
quantum computer architectures: the spin-based quantum dot QC, the
Cooper-pair-box-based QC, and the donor-nuclear-spin-based Si QC.

\section{Non-adiabatic Operations in a Double Quantum Dot}

Let us first discuss our work on the spin-based quantum dot quantum
computer (QDQC) in GaAs \cite{LD,HD}.  Here two-qubit operation is
based on the nearest neighbor exchange coupling, which produce the
exchange splitting between the ground singlet and triplet states. 
For small quantum dots, with large single particle
excitation energy $E_s$ and large on-site Coulomb repulsion energy
$E_C$ and at low temperatures ($k_B T \ll min\{E_s, E_C\}$), the low
energy dynamics is dominated by the electron spins.  In other
words, one can focus on the spin part of the two-electron Hilbert
space that involves only the ground singlet and triplet states and
cut off the rest of the Hilbert space.  For example, at $T \sim 100$
mK and with $min\{E_s, E_C\} \sim 1$ meV, the thermal occupation of
the higher energy orbital states is less than $10^{-50}$, which can be
safely neglected.  Thus one can quite faithfully prepare a single
electron in a single dot in its ground orbital state and/or two
electrons in a double dot in the ground singlet/triplet state
manifold.  For a double dot, after the state is initialized, as long as
the applied quantum gates satisfy adiabatic condition, the system
would remain in the ground state manifold, so that Heisenberg exchange
Hamiltonian would describe the dynamics of the double quantum dot
exactly.  However, the size of a gated quantum dot is limited from
below by gate and device dimensions, while the gate operating time is
limited from above by the electron spin decoherence time.  Thus it is
necessary to quantitatively assess the adiabatic condition for
two-qubit operations in a double dot of realistic dimensions, so as
to determine whether exchange gates can be sufficiently fast to
guarantee a large gate-time/decoherence-time ratio while slow
enough to produce correctably small non-adiabatic errors.  

We have performed a quantitative evaluation of the adiabatic
condition in a double quantum dot \cite{HD3} using the results
of our molecular orbital calculation of the double energy spectra
\cite{HD}.  Specifically, we prepare a two-electron state in the
ground singlet state with a high barrier between the double dot.
As the system evolves, the barrier height between the two dots
is first lowered, then raised back to the original value.  If
Heisenberg exchange Hamiltonian is exact for this system, its state
should remain in the ground singlet state.  Any loss from this state
would then constitute a leakage from the QC Hilbert space and a gate
error.

Our calculation is essentially an integration of the time-dependent
Schr\"{o}dinger equation for the two-electron double quantum dot: 
\begin{eqnarray}
\frac{\partial c_k(t)}{\partial t} & = & \sum_{i \neq k}^{N}
\frac{c_i(t)}{E_k(t) - E_i(t)} \langle k |\frac{\partial
H(t)}{\partial t}| i\rangle \,\nonumber  \\
& & \times \exp\left\{\frac{1}{i\hbar}
\int_{-\infty}^t (E_i(\tau) - E_k(\tau)) d\tau  \right\} \,.
\label{eq:coef}
\end{eqnarray}
Here $c_k(t)$ are the coefficients as we expand the two-electron
state on the instantaneous eigenstates $| k\rangle$:
$\psi(t) = \sum_k c_k(t) u_k(t)$ and $H(t) u_k(t) = E_k(t) u_k(t)$,
where $H(t)$ is the time-dependent system Hamiltonian.  The
explicit time-dependence of $H$ is in the inter-dot barrier height. 
Since initially the system is entirely in the ground singlet state,
$c_k(t=0) = \delta_{k0}$ for all $k$.  The energy spectra we
use are for a double dot with Gaussian confinements of 30 nm
radii and 40 nm inter-dot distance \cite{HD}.  The energy barrier
height $V_b$ ranges between 14 meV and 35 meV, corresponding to
exchange splitting of 280 $\mu$eV to 3.3 $\mu$eV.  By varying the
barrier variation time, we can quantitatively evaluate the change in
the ground singlet state population, thus obtaining a lower limit to
the gate operating time using the criterion of quantum error
correction code threshold. 

The result of our calculation is plotted in Fig.~1 \cite{HD3}.  The
leakage (y-axis) is defined as $1-|c_0|^2$ which is zero before the
gate is applied, and should be zero if the gate is perfectly
adiabatic.  Aside from several interesting features \cite{HD3},
Fig.~1 demonstrates that for gating time longer than $30 \sim 40$ ps,
leakage in our double dot system should be sufficiently small
($\lesssim 10^{-6}$) so that the currently available quantum error
correction schemes would be effective.  On the other hand, an
exchange splitting of 0.1 meV corresponds to about 20 ps gating time
for a swap gate \cite{LD} (with rectangular pulse) at the shortest. 
Therefore, adiabatic condition does not place an extra burden on the
operation of the two-qubit gates such as a swap---there is in general
no need to significantly increase the gating time in order to
accommodate the adiabatic requirement.

Notice that the current calculation is done for a pair of quite
small quantum dots.  Larger dots would have meant smaller excitation
energies and a threshold gating time that is longer in order to
satisfy the adiabatic condition.

\section{Non-sudden Operations in a Cooper Pair Box}
 
Another example we have considered is the Cooper pair box quantum
computer (CPBQC) \cite{HD3}.  The Hamiltonian of a Cooper pair
box (CPB) can be written on the basis of charge number states of the
box: 
\begin{eqnarray}
H & = & 4E_C (\hat{n}-n_g)^2 -E_J \cos\hat{\phi} \nonumber \\
& = & \sum_n \left[ 4E_C (n-n_g)^2 |n\rangle \langle n| \right. 
\nonumber \\
& & \left. -\frac{E_J}{2}(|n\rangle\langle n+1| + |n+1\rangle\langle n|)
\right],
\label{eq:CPB}
\end{eqnarray}
\begin{figure}[ht]
\centerline{
\epsfxsize=3.5in
\epsfbox{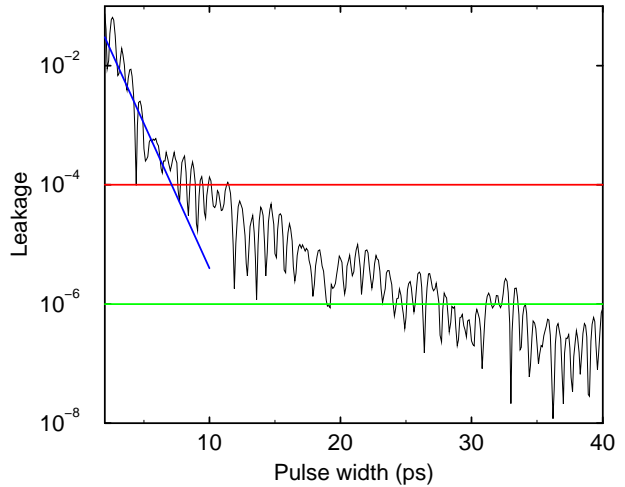}}
\protect\caption[Non-adiabatic leakage in a double QD]
{\sloppy{
Leakage as a function of the pulse width
$2\tau$ of the exchange gate.  The two-electron state is initially
in the ground singlet state.  The two horizontal lines
represents the commonly used thresholds for quantum error correction.
The fitted line at the small pulse-width indicates the initial
rapid decrease in the error rate (leakage) as the pulse becomes
wider (or the gate operation becomes slower).
}}
\label{fig-QD}
\end{figure}
\noindent where $E_C$ is the charging energy of a CPB, $E_J$
is the Josephson coupling between the CPB and an external
superconducting lead, $n_g$ represents the applied voltage on the
CPB in terms of an effective charge number, and $n$ refers to the
number of extra Cooper pairs in the box.  Due to the periodic
nature of the Josephson coupling, the eigenstates of a CPB form energy
bands.  The two states $|\!\uparrow\rangle$ and $|\!\downarrow\rangle$
for a CPB qubit correspond to the two lowest energy levels at
$n_g=1/2$, where the eigenstates are approximately
$|\!\downarrow\rangle=(|0\rangle + |1\rangle)/\sqrt{2}$ and
$|\!\uparrow\rangle=(|0\rangle - |1\rangle)/\sqrt{2}$ with a
splitting of about $E_J$.

Similar to the case of QDQC discussed above, higher excited states
play an important role in the dynamics of a CPBQC when it is
subjected to non-resonant operations \cite{Fazio}.  The particular
operation we considered is the sudden pulse gate to shift $n_g$, thus
bringing a system from a pure ground state ($|0\rangle$ at, e.g.
$n_g=1/4$) to a coherent superpositioned state ($(|\!\uparrow\rangle +
|\!\downarrow\rangle)/\sqrt{2}$ at $n_g=1/2$).  Such a
simple description of the pulse gate is only valid when $E_J/E_C
\rightarrow 0$.  Since $E_J$ determines the gate speed of a CPBQC,
such a condition is not practical for a realistic QC.  Furthermore,
in real experiments, the pulse gate always has finite rise/fall times
(non-sudden).  In Ref.~\cite{Nakamura}, the pulse rise time is
in the range of 30 to 40 ps.  Such gradual rise and fall of the
pulse gate inevitably lead to more errors, which have been considered
in the context of two-level systems \cite{Nakamura,Choi,Oh}.  What we
have done is to calculate the fidelity of the pulse gate taking into
account the finite rise/fall time, the higher excited states, and the
complete composition of all the eigenstates \cite{HD3}.    
\begin{figure}[ht]
\centerline{
\epsfxsize=3.5in
\epsfbox{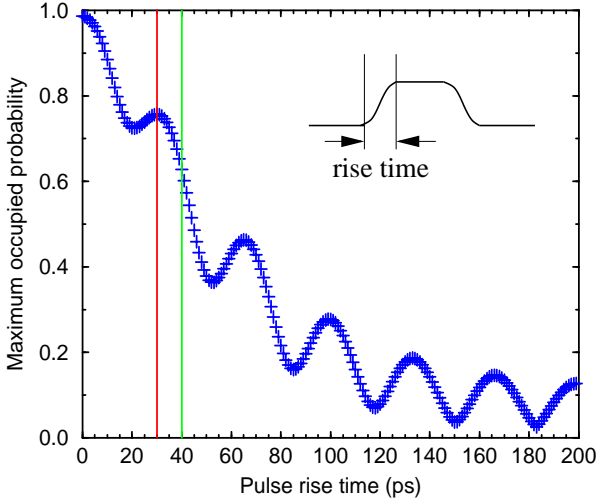}}
\protect\caption[Fidelity of a sudden pulse in a CPB] 
{\sloppy{    
State fidelity as a function of the finite rise/fall time of a pulse
gate in a single Cooper pair box.  The fidelity here is defined as
the maximum probability (which implies particular choices of pulse
duration $\tau_p$ as this probability varies periodically with
$\tau_p$) of the CPB in the first excited state after the
application of a pulse gate with the state starting from the ground
state (represented by $n_g$ goes from 0.255 to 0.5, then back to
0.255 after a period of time $\tau_p$).  The CPB is treated as a
multi-level system (on the basis of $|-10\rangle, \cdots,  |0\rangle,
|1\rangle, \cdots, |10\rangle$).  The lineshape of the rise/fall of
the pulse is a sinusoidal function of time.  The system
parameters are chosen as the values used in Ref.~\cite{Nakamura}.
The two vertical lines give the range of rise/fall time from the
same source.
}}
\label{fig-CPB}
\end{figure}

In Fig.~\ref{fig-CPB} we plot the state fidelity as a function of rise
time.  Here the state fidelity is defined as the maximum probability
for the CPB to be in the first excited state after the pulse gate
when $n_g$ returns to $0.255$.  Figure~\ref{fig-CPB} shows
that the fidelity of the pulse gate decreases oscillatorily instead
of monotonically as the pulse rise time increases.  The oscillations
(with periods around 30 ps) in the curves represent the coherent
evolution of the CPB during the rise/fall of the pulse voltage. 
For pulses used in Ref.~\cite{Nakamura} with rise/fall time in the
range of 30 to 40 ps, the fidelity is only 60 to 70 \%, apparently
not sufficient for manipulations required by quantum computation.
Further calculations also demonstrate that including higher excited
states is important in correctly evaluating the fidelity dependence
on the rise time of the non-resonant sudden pulse gate \cite{HD3}.

\section{Implications of Si Bandstructure}

The previous two examples demonstrate the interplay of non-resonant
gate operations and states from the full Hilbert space, and the
resulting leakage from the computational space.  Solid state
environment can affect the operation of a quantum computer in other
subtle ways.  For example, modern technology can produce extremely
pure silicon crystals which have the intrinsic property of very small
spin-orbit coupling.  Thus electron and nuclear spins in Si have a
very ``quiet'' environment---the spin relaxation times are extremely
long in Si \cite{Feher,Abragam}.  It is therefore natural to use Si
as a host material for spin-based quantum computer architectures
\cite{Kane}. However, Si is an indirect gap semiconductor.  There are
actually six equivalent minima in its conduction band that are away
from the center of the Fist Brillouin zone and close to the zone
boundary. The implication of this complexity is that confined
electron states (whether the confinement is provided by a donor or a
gate-produced electrostatic potential in the form of a quantum dot) in
general have contributing components from all the valleys, which can
then lead to atomic scale spatial oscillations of electronic
properties such as electron density and two-electron exchange
coupling.

We have performed a Heitler-London calculation for the two-electron
exchange splitting for two phosphorus donors in Si \cite{KHD}.  The
Si:P system is being studied as a candidate of nuclear spin
based quantum computers \cite{Kane,Clark}.  Donor electron exchange
is a crucial intermediary in the effective nuclear spin exchange
interaction that is the basis of the two-qubit operations in such a
quantum 
\begin{figure}[th]
\centerline{
\epsfxsize=3.5in
\epsfbox{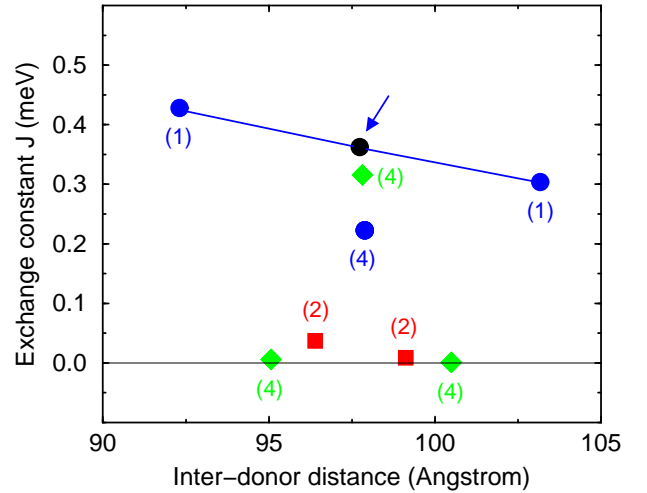}}
\protect\caption[Donor exchange in Si: effects of
hopping]
{\sloppy{Variations in the exchange coupling between two phosphorus
donors in Si.  The arrow points at the circle that represents the
value of electron exchange at the reference configuration with the
two donors   exactly along the [100] direction and separated by 18
lattice constants.   The circles connected by a line refer to pairs
along the [100] direction, displaced by  one lattice constant with
respect to the reference position. The rest of the symbols represent
displacements of one member of the donor pair into one of its first
(squares), second (diamonds), or third (circles) nearest neighbor
positions.  The numbers in the parenthesis next to the symbols are
their degeneracies, respectively.
}}
\label{fig-Si}
\end{figure}
\noindent computer architecture.  Our calculation indeed shows a
fast-varying exchange, as is demonstrated in Fig.~3, which shows that
a movement of one member of the donor pair into its nearest or second
nearest neighbor sites can completely suppress the 
exchange coupling between the two donor electrons.

In the original proposal of Si quantum computer \cite{Kane}, electron
exchange is tuned by applied gate voltages, which would shift the
electron wavefunctions.  Thus the two-qubit gates here are exposed to
the atomic-scale oscillations.  The direct implication of the
oscillatory exchange is that the gate voltages corresponding to the
peak exchange coupling have to be well-controlled, optimally close to
a local maximum where the exchange is least sensitive to the gate
voltage.  Since the oscillatory exchange period is close to lattice
spacing, the positioning of the donor electrons by the surface gates
must be controlled at least to that precision.

\section{Acknowledgment}

We thank financial support from ARDA and LPS
and collaborations (on the donor exchange in Si) and helpful
discussions with Belita Koiller.

\end{document}